\documentclass[prd,twocolumn,showpacs,preprintnumbers,amsmath,amssymb]{revtex4}
\usepackage{amsmath}
\usepackage{amsfonts}
\usepackage{amssymb}
\usepackage{graphicx}
\usepackage{dcolumn}% Align table columns on decimal point
\usepackage{bm}% bold math

\newcommand{\ima}{\hbox{Im}\,}
\newcommand{\rea}{\hbox{Re}\,}

\newcommand{\mr}{m_R^2}
\begin{document}

\title{Enhanced non-quark-antiquark and non-glueball $N_c$ behavior of 
light scalar mesons.}

\author{J. Nebreda}
\author{J. R. Pel\'aez}
\author{G. R\'{\i}os}
\affiliation{Departamento de F{\'\i}sica Te{\'o}rica II,
  Universidad Complutense de Madrid, 28040   Madrid,\ \ Spain}
  
\begin{abstract}
We show that the latest and very precise dispersive data 
analyses require a large and very unnatural fine-tuning 
of the $1/N_c$ expansion at $N_c=3$
if the $f_0(600)$ and $K(800)$ light scalar mesons
are to be considered predominantly $\bar qq$ states,
which is not needed for light vector mesons.
For this, we use scattering observables whose $1/N_c$ corrections are suppressed 
further than one power of $1/N_c$ for $\bar qq$ or glueball
states, thus enhancing contributions of other nature. 
This is achieved without using unitarized ChPT, but if it is used
we can also show that it is not just that the coefficients of the $1/N_c$ 
expansion are unnatural, but that the expansion itself does not
even follow the expected $1/N_c$ scaling of a glueball or a $\bar qq$ meson.
\end{abstract}

% PACS, the Physics and Astronomy
                             % Classification Scheme.
%\keywords{Suggested keywords}%Use showkeys class option if keyword
                              %display desired
\maketitle

%\section{Introduction}
Light scalar resonances play a relevant role for several fields of Physics: For
the nucleon-nucleon interaction, because they are largely responsible for the 
attractive part \cite{Johnson:1955zz} (with cosmological and anthropic implications).
for the QCD non-abelian nature, because 
some of these resonances have the  quantum numbers of the lightest glueball, 
also common to the vacuum and
hence of relevance for the spontaneous chiral symmetry breaking. 
Moreover, they are also of interest for the saturation \cite{Ecker:1988te} of the low energy constants of Chiral Perturbation Theory (ChPT) \cite{chpt}. However, the precise properties of these mesons, 
their nature, spectroscopic classification,
and even their existence---as for the $K(800)$ or $\kappa$---are still the object of
an intense debate. In particular, different models \cite{Jaffe:1976ig} suggest that they
may not be ordinary quark-antiquark mesons, but tetraquarks, 
meson molecules, glueballs, or a complicated mixture of all these.
The problem, of course, is that we do not know how to solve QCD at low energies.

However, since the QCD $1/N_c$ expansion is applicable at all energies, and
the mass and width $N_c$ dependence of $\bar{q}q$ mesons and glueballs 
is well known \cite{Witten:1980sp},
 the $N_c$ scaling of resonances becomes a powerful tool to classify them
and understand their nature.
In \cite{Pelaez:2003dy,Pelaez:2006nj} some of us studied the mass and width behavior of light resonances using 
ChPT---which is the QCD low energy effective Lagrangian---and unitarization with a dispersion relation.
It was found that the poles of the  $\rho(770)$ and $K^*(982)$  vectors behave predominantly 
as expected for $\bar{q}q$ states whereas those of the
$f_0(600)$, also called $\sigma$, and $K(800)$ scalars do not \cite{Pelaez:2003dy}.
Still, a possible subdominant $\bar{q}q$ component  for the $f_0(600)$
may arise naturally at two loops \cite{Pelaez:2006nj}
within ChPT (less so at one loop), but with a mass around 1 GeV or more. 

Of course, all these conclusions rely on unitarized ChPT and the assumption
that corrections, suppressed just by $1/N_c$, are of natural size.
Since $N_c$=3 in real life, this may not seem as a large suppression, even more 
when the meaning of ``natural size'' may not be clear for dimensional
parameters. 
%(for the $f_0(600)$, we might think about threshold or the half-width,
%$\sim 250-280$~MeV, the pole mass $\sim 450$~MeV, the ``peak mass'' $\sim 800$~MeV, etc...
%differing  by a factor of 3). 
For that reason, unitarized ChPT was useful to change $N_c$,
and reveal the $1/N_c$ scaling, no matter how unnatural the coefficients may appear. 

Here we will provide adimensional observables with corrections 
suppressed further than $1/N_c$, that can also be applied directly to real data at $N_c=3$,
without the need to extrapolate to larger $N_c$ using unitarized ChPT.

In particular,
resonances appearing in elastic two-body scattering
are commonly identified by three criteria.  
The $N_c$ behavior of one of these criteria---the associated 
pole in the unphysical sheet---was already studied in 
\cite{Pelaez:2003dy,Pelaez:2006nj}. 
A second possibility is to define 
the mass as the energy where the
phase shift reaches $\pi/2$, which both for $\pi\pi$ or $\pi K$ scattering
occurs relatively far from the $f_0(600)$ and $K(800)$ pole positions. 
This criterion 
was studied in 
\cite{Nieves:2009kh} for the $f_0(600)$ with a relatively 
inconclusive result about its assumed 
$\bar qq$ behavior. A more reliable parametrization and better data were called for and
we will provide them here together with more conclusive results. 
Third, the phase increases very fast in the resonance region
and the mass can be identified with
the maximum of the phase derivative. All three criteria roughly coincide for narrow resonances, but
the most physical definition is the latest, since it identifies
the resonance as a metastable state whose lifetime is the inverse of the width.
Note that this is the less evident feature 
both for the $f_0(600)$ and $K(800)$ and thus the phase derivative will 
become our preferred observable to test their $N_c$ dependence.

Let us then recall that partial waves generically scale as $1/N_c$, except at the 
resonance mass $m_R$. Actually, it has been found \cite{Nieves:2009kh} that 
if a resonance pole at $s_R=m_R^2-im_R \Gamma_R$
behaves as a $\bar qq$ \cite{Witten:1980sp}, i.e. $m_R\sim O(1)$ and $\Gamma_R\sim O(1/N_c)$,
then the phase shift satisfies
\footnote{Note our $t(s)$ has an opposite sign compared to that in \cite{Nieves:2009kh}.}:
 \begin{eqnarray}
 \delta(m_R^2)&=&\frac{\pi}{2}-\underbrace{\frac{\rea t^{-1}}{\sigma}
\Big\vert_{m_R^2}}_{O(N_c^{-1})}+O(N_c^{-3}),
 \label{deltaexpansion}\\
\delta'(m_R^2)&=&-\underbrace{\frac{(\rea t^{-1})'}{\sigma}
\Big\vert_{m_R^2}}_{O(N_c)}+O(N_c^{-1}), 
\label{derivativeexpansion}
 \end{eqnarray}
where $t(s)$ is the partial wave, $\sigma=2k/\sqrt{s}$
and $k$ is the meson center of mass momentum. Derivatives are taken with respect to $s$.
The $1/N_c$ counting of the different terms in the equations above 
comes from
the following expansions at $s=m_R^2$
\footnote{Note the corrected sign 
in front of $\sigma'$ which nevertheless does not affect
the results in \cite{Nieves:2009kh}.}:
\begin{eqnarray}
\rea t^{-1}&=&m_R\Gamma_R\Big[\frac{m_R\Gamma_R}{2}(\rea t^{-1})''-\sigma'\Big]+O(N_c^{-3}), \quad \label{retinvexpansion}\\
m_R\Gamma_R&=& \frac{\sigma}{(\rea t^{-1})'}+O(N_c^{-3}).\label{mRGR}
\end{eqnarray}
In brief, the corrections in Eqs.~\eqref{deltaexpansion} to \eqref{mRGR}
are suppressed by a further $1/N_c^2$ power
due to an expansion on the imaginary part of the pole, 
which scales like $\Gamma\sim 1/N_c$.
As nicely shown in \cite{Nieves:2009kh}, 
by expanding separately the real and imaginary parts 
of $t^{-1}$, only the $1/N_c^{2n+1}$ powers are kept 
on each expansion, leading to Eqs. \eqref{retinvexpansion} and \eqref{mRGR}.

Since we are interested in adimensional observables whose corrections are suppressed
further than just $1/N_c$,
we can recast Eqs.\eqref{deltaexpansion} and \eqref{derivativeexpansion}
%and \eqref{retinvexpansion}, 
as follows:
\begin{eqnarray}
\frac{\frac{\pi}{2}- \rea t^{-1}/\sigma}{\delta}\Big\vert_{m_R^2}&
\equiv&\Delta_1=1+\frac{a}{N_c^3},\label{adef}\\
-\frac{[\rea t^{-1}]'}{\delta'\sigma}\Big\vert_{m_R^2}&\equiv&\Delta_2=1+\frac{b}{N_c^2}.
\label{bdef}
%\\
%\frac{(\rea t^{-1})(\rea t^{-1})'}{\sigma[\frac{\sigma}{2}\frac{(\rea t^{-1})''}{(\rea t^{-1})'}-\sigma']}&\equiv&\Delta_3=1+\frac{c}{N_c^2},
\end{eqnarray}
Note that we have normalized each equation 
and extracted the leading $1/N_c$ dependence so that
the coefficients $a$ and $b$ should naturally be $O(1)$ or less.
It is relatively simple to make $a$ and $b$ much smaller than one 
with cancellations with natural higher order $1/N_c$ contributions,
but very unnatural to make them much larger.

Now, in Table~\ref{tab:Nc3results} we show the resulting $a$ and $b$ 
for the lightest resonances found in $\pi\pi$ and $\pi K$ elastic scattering.
Before describing in  detail the calculations, let us observe
that {\it for the $\rho(770)$ and $K^*(892)$ vector resonances
 all parameters
are of order one or less, as expected for
$\bar qq$ states}.   
In contrast, {\it for the $f_0(600)$ and $K(800)$ scalar resonances we 
find that all parameters are larger, by two orders of magnitude,
than expected for $\bar qq$ states}. 
This is one of the main results of this work and make 
the $\bar qq$ interpretation of both scalars extremely unnatural. 
\begin{table}
  \centering
  \begin{tabular}{crrrr}
    & $\rho(770)$ & $K^*(892)$ & $f_0(600)$ & $K(800)$ \\\hline
\rule[-1mm]{0mm}{5mm}$a$    &    $-0.06\pm0.01$  &  0.02  &  $-252^{+119}_{-156}$ & -2527 \\
\rule[-1mm]{0mm}{5mm}$b$    &  $0.37 ^{+0.04}_{-0.05}$    & 0.16 &$77^{+28}_{-24}$ & 162 \\
%\rule[-1mm]{0mm}{5mm}$c$  &$-0.9\pm1.2$ & & $75^{+34}_{-26}$ & \\
\hline
  \end{tabular}
  \caption{Normalized coefficients of the $1/N_c$ expansion for different resonances.
For $\bar qq$ resonances, all them are expected to be of order one or less.}
  \label{tab:Nc3results}
\end{table}

Let us now describe in detail our calculations and their
different degree of precision and reliability. 
As commented above, the $f_0(600)$  ``Breit-Wigner'' mass was already
studied \cite{Nieves:2009kh} using Eq. \eqref{deltaexpansion},
but no conclusion was reached
on whether the deviations were consistent
with the $1/N_c$ suppression or not. This was partly attributed 
to the limited  reliability
of the conformal parametrization or
unitarized ChPT---whose phase never reaches $\pi/2$-- used in \cite{Nieves:2009kh}. To overcome this
caveat we are now using the recent, very precise and reliable 
{\it output} of the data analysis in 
\cite{GarciaMartin:2011cn} constrained to satisfy
 once subtracted coupled dispersion relations---or GKPY equations---as well as Roy equations, which is therefore model independent and 
specially suited to obtain the $f_0(600)$ pole \cite{nosotros}.
Note that this analysis incorporates the very recent and reliable data on $K_{l4}$
decays from NA48/2 \cite{Batley:2010zza,nosotros}, 
which is a key factor in attaining high levels of precision.
The analysis in \cite{GarciaMartin:2011cn} is also
in good agreement with previous dispersive results
based on standard Roy equations \cite{CGL}.
We have followed the same rigorous approach 
for the $\rho(770)$, although, being so narrow, the conformal 
unconstrained data analysis and the IAM yield very similar results.
The uncertainties we quote for both the $f_0(600)$ and $\rho(770)$
cover the uncertainties in the output of the dispersive representation.

In this work we also deal with strange resonances in $\pi K$ scattering.
For the scalar $K(800)$ we have also 
used a rigorous dispersive calculation, namely, that in \cite{DescotesGenon:2006uk},
which uses 
Roy-Steiner equations to determine the isospin 1/2 scalar channel 
of $\pi K$ scattering, although this time 
we can only provide a central value.
Note, however, that the value of $m_R^2$ 
obtained in that analysis 
is located below threshold, so that the phase shift is ill defined at $m_R^2$. 
Nevertheless, we have been using the $m_R$ mass definition to allow
for an easier comparison with \cite{Nieves:2009kh}, 
but the definition $\sqrt{s_R}=m-i\Gamma/2$
is equally valid and is actually the standard choice used in the context
of scalar mesons.
Moreover, the $N_c$ scaling of
Eqs.~\eqref{deltaexpansion} and \eqref{derivativeexpansion}
does not change if we evaluate the quantities at $s=m^2$, instead of $m_R^2$,
since $m^2$ differs from 
$m_R^2$ in $\Gamma^2/4$, which is $O(N_c^{-2})$. 
Thus, the values for the $K(800)$ 
in Table~\ref{tab:Nc3results} correspond to this choice.
For the vector $K^*(892)$ there are no very precise 
purely dispersive descriptions of the existing data and we therefore rely 
on a single partial wave dispersion relation and SU(3) ChPT to one-loop
to determine its subtraction constants 
(this is known as ChPT unitarized with the single channel
Inverse Amplitude Method (IAM) \cite{IAM}), 
which we will briefly explain in the next section. 
We have applied the same method to the $\rho(770)$ and 
the results lie within 50\% of their central value when using the GKPY dispersive
representation. Since the $K^*(892)$ is narrower 
than the $\rho(770)$, the IAM is likely to provide a 
better approximation than in the $\rho(770)$ case, 
but even with that 50\% uncertainty, it is enough
to check that the $a$ and $b$ parameters are smaller than one.

% \begin{table}[h]
%   \centering
%   \begin{tabular}{c|cccc}
% \rule[0.3cm]{0.cm}{0.cm}& \hspace{.2cm}  $\pi/2$ \hspace{.2cm} & \hspace{.2cm}  $|\delta'\,\Delta|$ \hspace{.2cm} & \hspace{.2cm} $\frac{\delta-\pi/2}{\delta'\Delta}-1$ \hspace{.2cm} & $|\delta-\pi/2-\delta'\Delta|$\\
% \rule[0.35cm]{0.cm}{0.cm}& $O(1)$  &  $O(N_c^{-1})$     & $O(N_c^{-2})$                   &  $O(N_c^{-3})$                \\
% \rule[0.3cm]{0.cm}{0.cm}$\simeq$& 1       &   0.33             & 0.11                            & 0.037                         \\\hline
% \rule[0.3cm]{0.cm}{0.cm}$\rho(770)$& 1.57    &   0.19             & 0.06                            & 0.011                         \\
% \rule[0.3cm]{0.cm}{0.cm}$K^*(892)$& 1.57    &             &                 &                        \\
% \rule[0.3cm]{0.cm}{0.cm}$f_0(600)$& 1.57    &   0.0047           & 280                             & 1.30                          \\
% \rule[0.3cm]{0.cm}{0.cm}$K(800)$& 1.57    &             &                  &                         \\
%   \end{tabular}
%   \caption{Hierarchy of terms evaluated at $s=m_R^2$}
% \label{tab:hierarchy}
% \end{table}

There is, of course, another way of interpreting our results, 
which is that due to the large $1/N_c$ coefficients of
the $f_0(600)$ the series simply does not converge.
In particular,
Eq.\eqref{deltaexpansion}, which was thoroughly considered in \cite{Nieves:2009kh},
 is obtained as an expansion of  
${\rm arctan}(x)= x-x^3/3..$. In this way we could explain why the 
$a=-0.06\pm0.01$ coefficient is so small for the $\rho(770)$:
it is simply the effect of calculating $a=\tilde a^3/3$ with $\tilde a= 0.56^{+0.03}_{-0.04}$, which is now naturally of $O(1)$. We could try the same procedure for the $f_0(600)$,
assuming its series expansion is that of a $\bar qq$,
to find $\tilde a= 9.1^{}_{}$, still rather unnatural, but of course, this value
 makes no sense since the whole 
series would not be converging and terms higher than $1/N_c^3$
would become dominant.

This is one of the reasons why despite being only suppressed by $1/N_c^2$ 
instead of $1/N_c^3$, we also provide the expansion in  Eq.\eqref{bdef}
obtained from the derivative of the amplitude. In this case the $b/N_c^2$ term is not the square of a natural $1/N_c$ quantity, i.e.,
\begin{equation}
\frac{b}{N_c^2}=\frac{\rea t^{-1}}{\sigma}
\Big[\frac{\sigma'}{(\rea t^{-1})'}-\frac{\rea t^{-1}}{\sigma}\Big]+O(N_c^{-4}).
\end{equation}
Despite containing a cancellation between two $1/N_c$ terms,
its value for the $\rho(770)$ is rather natural. However, once again, the 
value for the scalars is almost two orders of magnitude larger than expected.

In the previous analysis it is very relevant that the width of the resonance
is suppressed with additional $1/N_c$ powers with respect to the mass.
Actually, it is rather straightforward to extend 
the formalism to study the assumption that the 
$f_0$(600) could be predominantly a glueball, since then
$m_R\sim O(1)$ and $\Gamma_R\sim O(1/N_c^2)$ 
\cite{Witten:1980sp,LlanesEstrada:2011kz}. As a consequence, 
for the glueball case, the scaling of Eqs.~\eqref{retinvexpansion} and~\eqref{mRGR} changes and so does that of $\delta(m_R^2)$ and $\delta'(m_R^2)$:
 \begin{eqnarray}
 \delta(m_R^2)&=&\frac{\pi}{2}-\underbrace{\frac{\rea t^{-1}}{\sigma}
\Big\vert_{m_R^2}}_{O(N_c^{-2})}+O(N_c^{-6}),
 \label{deltaexpansionglueball}\\
\delta'(m_R^2)&=&\underbrace{\frac{(\rea t^{-1})'}{\sigma}
\Big\vert_{m_R^2}}_{O(N_c^2)}+O(N_c^{-2}).
\label{derivativeexpansionglueball}
 \end{eqnarray}
Much as it was done in Eqs.~\eqref{adef} and \eqref{bdef}, 
in order to make explicit this further $N_c$ suppression we can define 
some new parameters $a'$ and $b'$ that 
should be of $O(1)$ if the resonance was a glueball:
\begin{equation}
\Delta_1=1+\frac{a'}{N_c^6},\qquad
\Delta_2=1+\frac{b'}{N_c^4}.\label{abpdef}
\end{equation}
Following the same procedure as before we obtain for the $f_0(600)$,
$a'=-6800^{+3200}_{-4200}$ and $b'=2080^{+760}_{-650}$.
In other words, a very dominant or pure glueball nature for the 
$f_0(600)$ is very disfavored by the $1/N_c$ expansion,
even more than the $\bar qq$ interpretation. This is because
it would require even more unnatural coefficients, this time  too large by
three to four orders of magnitude.

Of course, as we did for the $\bar qq$ case, we could worry 
about the fact that,
due to the ${\rm arctan}(x)= x-x^3/3+...$ expansion, the $a'$ should have been
interpreted as
 $a'=\tilde a'^3/3$. But even with that interpretation we would still
find $\tilde a'=27^{+5}_{-7}$, again rather unnatural. Once more,  and as it happened in the $\bar qq$ case,
the $b'$ parameter does not correspond 
to the fourth power of any natural quantity, so that its value is genuinely unnatural, disfavoring the glueball interpretation.

Let us remark that in the case of 
tetraquarks or molecules, the width is not expected to be suppressed
with additional $1/N_c$ powers with respect to the mass of the resonance
\cite{Jaffe,LlanesEstrada:2011kz}. Thus, our previous formalism does not apply.
Furthermore, it is most likely that scalars are a mixture of different components.
Therefore our results, while showing that neither the $\bar qq$ or a  glueball
are favored as dominant components of light scalars, do not exclude that these
structures could be mixed with other components that would 
dominate the $1/N_c$ expansion with a different $N_c$ behavior \footnote{
For some preliminary attempts at disentangling the composition of light scalars
using their leading $1/N_c$ behavior we refer the reader to \cite{LlanesEstrada:2011kz}.}.

In summary, we have just shown that if, for the light
scalar mesons, we study $\bar qq$ or glueball $1/N_c$ expansions as those in
Eqs.~\eqref{adef}, \eqref{bdef} and \eqref{abpdef},
their coefficients come out very unnatural, suggesting
that these resonances cannot be described as 
{\it predominantly} made of a quark and an antiquark or a glueball.
Note that, contrary to our previous 
works \cite{Pelaez:2003dy,Pelaez:2006nj},  
this conclusion has been reached
{\it from dispersive analyses of data}, 
without extrapolating to $N_c\neq3$ using unitarized ChPT.

However, unitarized ChPT will be used next to
calculate the $\Delta_i-1$ observables, in order to show
that, for scalars, what really happens is that
they do not even follow the $1/N_c$ expansion
of $\bar qq$ or glueball states given in 
Eqs.~\eqref{adef}, \eqref{bdef}  and \eqref{abpdef}, thus explaining the
need for unnatural coefficients if a $\bar qq$ or glueball-like
expansion is assumed.

{\bf The Inverse Amplitude Method:}
The elastic IAM \cite{IAM} uses ChPT to evaluate the subtraction 
constants and the left cut of a dispersion relation for the inverse
of the partial wave. The elastic right cut is exact, since the
elastic unitarity condition $\ima t=\sigma |t|^2$,
fixes $\ima t^{-1}=-\sigma$. Note that the IAM is 
derived only from elastic 
unitarity, analyticity in the form of a dispersion
relation, and ChPT, which is only used at low energies. It 
satisfies exact elastic unitarity and reproduces meson-meson
scattering data up to energies $\sim$ 1 GeV. It can be 
analytically continued into the second Riemann sheet where poles
associated to resonances are found. In particular, we find the
$\rho(770)$ and $f_0(600)$, 
as well as the $K^{*}(892)$ and the $K(800)$
 resonances as poles in $\pi\pi$ and $\pi K$ scattering amplitudes, 
respectively.

The dependence 
on the QCD number of colors, $N_c$, 
is implemented \cite{Pelaez:2003dy,Pelaez:2006nj} through the
leading $N_c$ scaling of the ChPT low energy constants (LECs), 
and is 
model independent \cite{chpt,chptlargen,Pelaez:2006nj}. 
Fortunately, for Eqs.\eqref{adef} and \eqref{bdef} to hold, only the leading $1/N_c$ behavior is needed.
Note also that the IAM does not have any other
parameters where uncontrolled $N_c$ dependence could hide---as it happens in other unitarization methods---so that the IAM allows us to check the scaling of the $\Delta_i-1$
in Eqs.\eqref{adef} and \eqref{bdef}.

{\it The $SU(2)$ IAM:} Only the 
non-strange $f_0(600)$ and $\rho(770)$ resonances can be checked, 
but we can do it unitarizing with the IAM the corresponding
 partial waves either to one or two loops.
We simply scale
$f_{N_c}\rightarrow f \sqrt{N_c/3}$, the one loop constants as
$l^r_{i,\,N_c}\rightarrow l_i^r N_c/3$ and 
the two loop ones as $r_{i,\,N_c}\rightarrow r_i (N_c/3)^2$.

\begin{figure*}
  \centering
  \vbox{
    \includegraphics[scale=1]{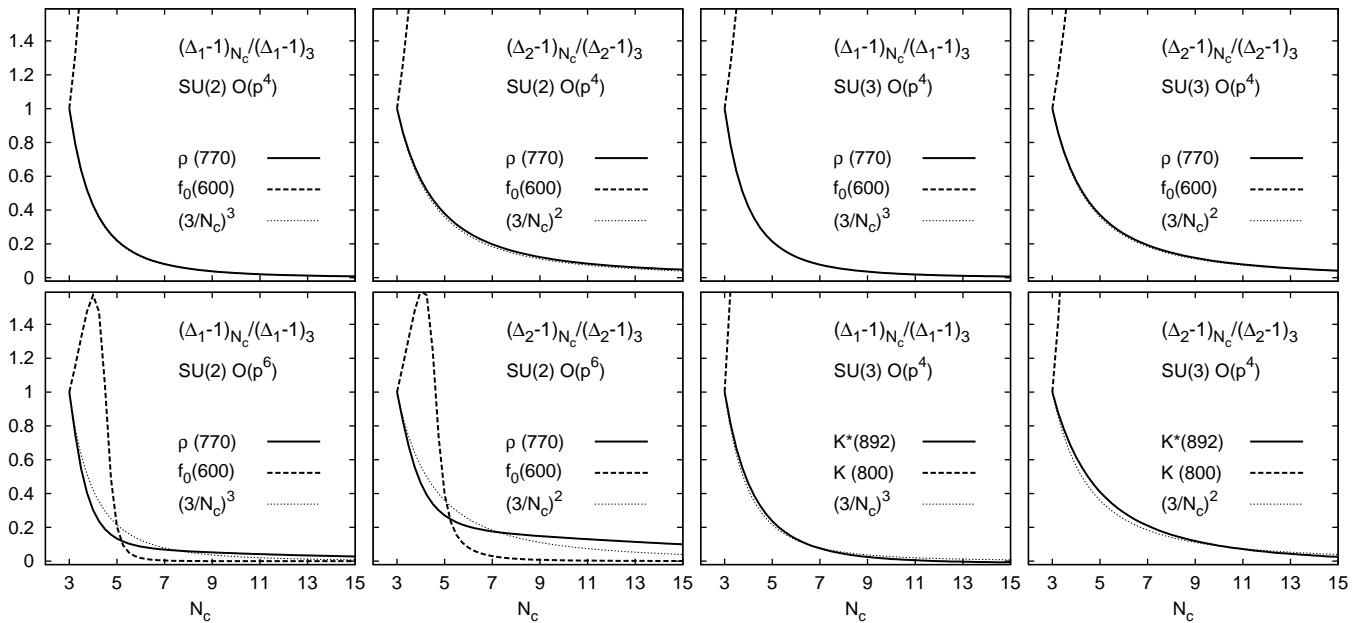}
  }
  \caption{$1/N_c$ scaling of the $\Delta_i-1$ observables 
    normalized to their $N_c=3$ value for light scalar and vector mesons, 
    using unitarized ChPT within $SU(2)$ or $SU(3)$ and to one or two loops---$O(p^4)$ 
    and $O(p^6)$, respectively. 
\label{figureab}}
\end{figure*}

Thus, in the two first columns of Fig.~\ref{figureab} we show, for the $\rho(770)$ and $f_0(600)$ resonances,  the scaling 
of the $\Delta_i-1$ 
 both to one loop (upper panels) 
and two loops (lower panels). Note we have normalized them
to their $N_c=3$ value, in order to cancel the leading part of
the $a$ and $b$ coefficients and thus extract the 
leading $1/N_c^k$ behavior of Eqs.~\eqref{adef} and \eqref{bdef}.
For the one-loop calculations we use the set of LECs in \cite{Nebreda:2011di},
whereas for the two-loop calculation we use the fit D from 
\cite{Pelaez:2010fj,Nebreda:2011di}. We have checked that similar 
results are obtained when using other sets of LECs in these references or
the estimates from resonance saturation \cite{Ecker:1988te}. 

We can observe that the scaling for the $\rho(770)$ observables
overlaps with the expectation for the leading behavior of $\bar{q}q$ states. 
However, in the case of the $f_0(600)$ the scaling is completely 
different. To one loop the $f_0(600)$
observables grow instead of decreasing. Let us note however, 
that, for $N_c$ larger than $\sim 10$, the $f_0(600)$ pole lies on the third
quadrant of the complex plane. Before that happens, the value of $\mr$
becomes less than $4 m_\pi^2$ and the
phase shift has no physical meaning so that Eqs.~\eqref{adef} and \eqref{bdef}
do not hold. 
This behavior does not occur to two loops. Actually we find again
the $f_0(600)$ behavior 
already observed in \cite{Pelaez:2006nj}, where, for $N_c$ close to 3, 
the width grows as in the one-loop case (and so do the observables here),
but for larger $N_c$ the $f_0(600)$ starts behaving more as a $\bar qq$.
Note that this $\bar qq$ behavior appears at a mass somewhat bigger than 1 GeV. 
This was a hint of the $f_0(600)$ being  a mixture of a
 predominantly non-$\bar qq$ component and, at least, a 
subdominant $\bar{q}q$ component with a mass much heavier that the physical one,
which is the one that survives at large $N_c$.
In terms of the $\Delta_i-1$ observables defined here, this translates into a 
growth close to $N_c=3$ and a decrease at larger $N_c$.
Therefore, it is not only that the $a$ and $b$ coefficients of the $f_0(600)$ are
too large as shown in the previous section, but that the scaling itself does
not correspond to a $\bar qq$ state (and even less so to a glueball).
To two-loops, the $\rho(770)$ does not
follow exactly the leading behavior of $\bar qq$ states but decreases slightly faster,
which can be naturally explained due 
to subleading effects or to a possible 
small pion cloud contribution.

{\it The $SU(3)$ IAM:} 
Now we can study the scaling of $\Delta_i-1$
not only for the $\rho(770)$ and $f_0(600)$, 
but also for the $K^*(892)$ and $K(800)$ resonances, although in this case
the elastic unitarized amplitudes are available only to one loop 
\cite{GomezNicola:2001as,Nebreda:2010wv}.
We have now eight LECs, called $L_i(\mu)$, that scale \cite{chpt,chptlargen} 
as $L_{i,\,N_c}\rightarrow L_i (N_c/3)$ for 
$i=2,3,5,8$, while $2L_1-L_2$, $L_4$, $L_6$ and $L_7$ do not change with $N_c$.

In the third and fourth columns of Fig.~\ref{figureab} 
we show the results found using the set of LECs called Fit~II in \cite{Nebreda:2010wv}.
Similar results are obtained with Fit~I or the estimates from
resonance saturation in \cite{Ecker:1988te}.
In the upper panels we simply reobtain within the $SU(3)$ formalism the same
results we obtained for the $\rho(770)$ and $f_0(600)$ within the $SU(2)$
formalism to one loop. In the lower panels we show the results for the 
light vector $K^*(892)$, following nicely the $\bar qq$ expectations,  as
well as the results for the scalar $K(800)$, which has a very similar behavior
to the $f_0(600)$, at odds with a dominant $\bar qq$ or glueball nature.

{\bf Summary:} In this work we have studied the $1/N_c$ expansion of
the meson-meson scattering phase-shifts around the pole mass of a $\bar qq$ 
or glueball resonance.
In particular we have defined observables whose corrections are suppressed 
further than just one power of $N_c$, paying particular attention to the
derivative of the phase, which provides a physical and intuitive definition
of a resonance. Using recent and very precise dispersive 
data analyses we have shown that  
if we assume a $\bar qq$ or glueball behavior for the $f_0(600)$ and $K(800)$, the coefficients of the
expansion of such observables turn out unnaturally large. This is shown without
using ChPT or extrapolating beyond $N_c=3$. 
Moreover, 
when using unitarized ChPT, we have shown that it is 
the very $1/N_c$ scaling of the 
observables which does not follow the pattern of the 
$1/N_c$ expansion expected for $\bar qq$ or glueball states.

{\it Acknoledgments.} We thank J. Nieves and E. Ruiz-Arriola for useful discussions, checks and suggestions and B. Mussallam for the results of his Roy-Steiner dispersive analysis of the $K(800)$ channel.

%which is also reproduced in the $\pi K$ channel, where the vector resonance $K^*$ scales as a 
%$\bar{q}q$ and the scalar $\kappa$ behaves much like the $f_0(600)$. 

%\begin{figure}
%  \centering
%  \vbox{
%    \includegraphics[scale=1.45]{obs2-SU3.ps} 
%    }
%  \caption{$N_c$ scaling of the observable \eqref{obs} in one-loop SU(3) normalized 
%  its physical value for the $\rho$ and $f_0(600)$ resonances (upper panel) and for 
%  the $K^*$ and the $\kappa$ (lower panel).}
%  \label{obs-SU3}
%\end{figure}


\begin{thebibliography}{99}

\footnotesize


\bibitem{Johnson:1955zz}
  M.~H.~Johnson and E.~Teller,
  %``Classical Field Theory of Nuclear Forces,''
  Phys.\ Rev.\  {\bf 98}, 783 (1955).
  %%CITATION = PHRVA,98,783;%%


%\cite{Ecker:1988te}
\bibitem{Ecker:1988te}
  G.~Ecker, J.~Gasser, A.~Pich and E.~de Rafael,
  %``The Role Of Resonances In Chiral Perturbation Theory,''
  Nucl.\ Phys.\  B {\bf 321}, 311 (1989).
  %%CITATION = NUPHA,B321,311;%%
%\cite{Donoghue:1988ed}
%\bibitem{Donoghue:1988ed}
  J.~F.~Donoghue, C.~Ramirez and G.~Valencia,
  %``The Spectrum of QCD and Chiral Lagrangians of the Strong and Weak
  %Interactions,''
  Phys.\ Rev.\  D {\bf 39}, 1947 (1989).
  %%CITATION = PHRVA,D39,1947;%%


\bibitem{chpt}
S. Weinberg, Physica {\bf A96} (1979) 327.
%\cite{Gasser:1983yg}
%\bibitem{Gasser:1983yg}
J.~Gasser and H.~Leutwyler,
%``Chiral Perturbation Theory To One Loop,''
Annals Phys.\  {\bf 158} (1984) 142;
%%CITATION = APNYA,158,142;%%
%\cite{Gasser:1984gg}
%\bibitem{Gasser:1984gg}
%J.~Gasser and H.~Leutwyler,
%``Chiral Perturbation Theory: Expansions In The Mass Of The Strange Quark,''
Nucl.\ Phys.\ B {\bf 250} (1985) 465.
%%CITATION = NUPHA,B250,465;%%

%\cite{Jaffe:1976ig}
\bibitem{Jaffe:1976ig}
  R.~L.~Jaffe,
  %``Multi-Quark Hadrons. 1. The Phenomenology Of (2 Quark 2 Anti-Quark)
  %Mesons,''
  Phys.\ Rev.\  D {\bf 15}, 267 (1977);
  %%CITATION = PHRVA,D15,267;%%
%\cite{Jaffe:2007id}
%\bibitem{Jaffe:2007id}
%  R.~L.~Jaffe,
  %``Ordinary and extraordinary hadrons,''
  AIP Conf.\ Proc.\  {\bf 964}, 1 (2007)
  [Prog.\ Theor.\ Phys.\ Suppl.\  {\bf 168}, 127 (2007)].
%  [arXiv:hep-ph/0701038].
  %%CITATION = PTPSA,168,127;%%
%
%
%\bibitem{jaffenew1}
  R.~L.~Jaffe and F.~E.~Low,
  %``The Connection Between Quark Model Eigenstates And Low-Energy Scattering,''
  Phys.\ Rev.\  D {\bf 19}, 2105 (1979).
  %%CITATION = PHRVA,D19,2105;%% 
%
%
%\bibitem{isgur1}
  J.~D.~Weinstein and N.~Isgur,
  %``Do Multi-Quark Hadrons Exist?,''
  Phys.\ Rev.\ Lett.\  {\bf 48}, 659 (1982);
  %%CITATION = PRLTA,48,659;%%
%
%
%\bibitem{isgur2}
%  J.~D.~Weinstein and N.~Isgur,
  %``The Q Q Anti-Q Anti-Q System In A Potential Model,''
  Phys.\ Rev.\  D {\bf 27}, 588 (1983);
  %%CITATION = PHRVA,D27,588;%%
%
%
%\bibitem{isgur3}
%  J.~D.~Weinstein and N.~Isgur,
  %``K anti-K Molecules,''
  Phys.\ Rev.\  D {\bf 41}, 2236 (1990).
  %%CITATION = PHRVA,D41,2236;%%
%
%
%\bibitem{janssen}
  G.~Janssen, B.~C.~Pearce, K.~Holinde and J.~Speth,
  %``On the structure of the scalar mesons f0 (975) and a0 (980),''
  Phys.\ Rev.\  D {\bf 52}, 2690 (1995).
%  [arXiv:nucl-th/9411021].
  %%CITATION = PHRVA,D52,2690;%%
%
%
%\cite{Achasov:1997ih}
%\bibitem{Achasov:1997ih}
  N.~N.~Achasov and V.~V.~Gubin,
  %``Searches for scalar a0 and f0 mesons in the reactions  e+ e- --> gamma pi0
  %pi0 (eta),''
  Phys.\ Rev.\  D {\bf 56} (1997) 4084
%  [arXiv:hep-ph/9703367].
  %%CITATION = PHRVA,D56,4084;%%
%\cite{Achasov:2001cj}
%\bibitem{Achasov:2001cj}
%  N.~N.~Achasov and V.~V.~Gubin,
  %``Interference pattern analysis in the decays Phi --> gamma pi eta and Phi
  %--> gamma pi0 pi0,''
  Phys.\ Atom.\ Nucl.\  {\bf 65}, 1528 (2002)
  [Yad.\ Fiz.\  {\bf 65}, 1566 (2002\ PHRVA,D63,094007.2001)].
%  [arXiv:hep-ph/0101024].
  %%CITATION = PHRVA,D63,094007;%%
%
%
%\cite{Minkowski:1998mf}
%\bibitem{Minkowski:1998mf}
  P.~Minkowski and W.~Ochs,
  %``Identification of the glueballs and the scalar meson nonet of lowest
  %mass,''
  Eur.\ Phys.\ J.\  C {\bf 9}, 283 (1999).
%  [arXiv:hep-ph/9811518].
  %%CITATION = EPHJA,C9,283;%%
%
%
%\bibitem{Oller:1997ti}
  J.~A.~Oller and E.~Oset,
  %``Chiral symmetry amplitudes in the S-wave isoscalar and isovector  channels
  %and the sigma, f0(980), a0(980) scalar mesons,''
  Nucl.\ Phys.\  A {\bf 620}, 438 (1997)
  [Erratum-ibid.\  A {\bf 652}, 407 (1999)].
%  [arXiv:hep-ph/9702314].
  %%CITATION = NUPHA,A620,438;%%
%
%
%\bibitem{eef}
E.~van~Beveren and G.~Rupp,
  Eur.\ Phys.\ J.\  C {\bf 22}, 493 (2001).
%
%
%\cite{Vijande:2005jd}
%\bibitem{Vijande:2005jd}
  J.~Vijande, A.~Valcarce, F.~Fernandez and B.~Silvestre-Brac,
  %``Nature of the light scalar mesons,''
  Phys.\ Rev.\  D {\bf 72} (2005) 034025.
%  [arXiv:hep-ph/0508142].
  %%CITATION = PHRVA,D72,034025;%%
%
%
%\cite{Hyodo:2010jp}
%\bibitem{Hyodo:2010jp}
  T.~Hyodo, D.~Jido and T.~Kunihiro,
  %``Nature of the sigma meson as revealed by its softening process,''
  Nucl.\ Phys.\  A {\bf 848} (2010) 341
%  [arXiv:1007.1718 [hep-ph]].
  %%CITATION = NUPHA,A848,341;%%
%
%
%\cite{Kaminski:2009qg}
%\bibitem{Kaminski:2009qg}
  R.~Kaminski, G.~Mennessier, S.~Narison,
  %``Gluonium nature of the sigma/f(0)(600) from its coupling to K anti-K,''
  Phys.\ Lett.\  {\bf B680}, 148-153 (2009).
%  [arXiv:0904.2555 [hep-ph]].
%\cite{Forkel:2010gu}
%\bibitem{Forkel:2010gu}
  H.~Forkel,
  %``Light scalar tetraquarks from a holographic perspective,''
  Phys.\ Lett.\  {\bf B694}, 252-257 (2010).
%  [arXiv:1007.4341 [hep-ph]].



\bibitem{Witten:1980sp}
%\cite{'tHooft:1973jz}
%\bibitem{'tHooft:1973jz}
G.~'t Hooft,
%``A Planar Diagram Theory For Strong Interactions,''
Nucl.\ Phys.\ B {\bf 72} (1974) 461.
%%CITATION = NUPHA,B72,461;%%
%\cite{Witten:1980sp}
%\bibitem{Witten:1980sp}
E.~Witten,
%``Large N Chiral Dynamics,''
Annals Phys.\  {\bf 128} (1980) 363.
%%CITATION = APNYA,128,363;%%

\bibitem{Pelaez:2003dy}
  J.~R.~Pelaez,
  %``On the nature of light scalar mesons from their large $N_c$ behavior,''
  Phys.\ Rev.\ Lett.\  {\bf 92}, 102001 (2004).

\bibitem{Pelaez:2006nj}
  J.~R.~Pelaez and G.~Rios,
  %``Nature of the f_0(600) from its N_c dependence at two loops in unitarized
  %Chiral Perturbation Theory,''
  Phys.\ Rev.\ Lett.\  {\bf 97}, 242002 (2006).
  
\bibitem{Nieves:2009kh}
  J.~Nieves and E.~Ruiz Arriola,
  %``Meson Resonances at large Nc: Complex Poles vs Breit-Wigner Masses,''
  Phys.\ Lett.\  B {\bf 679}, 449 (2009).
 
%\cite{GarciaMartin:2011cn}
\bibitem{GarciaMartin:2011cn}
  R.~Garcia-Martin, R.~Kaminski, J.~R.~Pelaez, J.~Ruiz de Elvira and F.~J.~Yndurain,
  %``The pion-pion scattering amplitude. IV: Improved analysis with once
  %subtracted Roy-like equations up to 1100 MeV,''
  Phys.\ Rev.\  D {\bf 83} (2011) 074004
%  [arXiv:1102.2183 [hep-ph]].
  %%CITATION = PHRVA,D83,074004;%%


\bibitem{nosotros}
%\cite{GarciaMartin:2011jx}
%\bibitem{GarciaMartin:2011jx}
  R.~Garcia-Martin, R.~Kaminski, J.~R.~Pelaez, J.~Ruiz de Elvira,
  %``Precise determination of the f0(600) and f0(980) pole parameters from a dispersive data analysis,''
  Phys.\ Rev.\ Lett.\  {\bf 107}, 072001 (2011).
%  [arXiv:1107.1635 [hep-ph]].


\bibitem{Batley:2010zza}
%\bibitem{Batley:2010zza}
  J.~R.~Batley {\it et al.} [ NA48-2 Collaboration ],
  %``Precise tests of low energy QCD from K(e4)decay properties,''
  Eur.\ Phys.\ J.\  {\bf C70}, 635-657 (2010).
  
%\cite{Colangelo:2001df}
\bibitem{CGL}
  G.~Colangelo, J.~Gasser, H.~Leutwyler,
  %``pi pi scattering,''
  Nucl.\ Phys.\  {\bf B603}, 125-179 (2001).
%  [hep-ph/0103088].
%\cite{Ananthanarayan:2000ht}
%\bibitem{Ananthanarayan:2000ht}
  B.~Ananthanarayan, G.~Colangelo, J.~Gasser, H.~Leutwyler,
  %``Roy equation analysis of pi pi scattering,''
  Phys.\ Rept.\  {\bf 353}, 207-279 (2001).
%  [hep-ph/0005297].

%\cite{DescotesGenon:2006uk}
\bibitem{DescotesGenon:2006uk}
  S.~Descotes-Genon, B.~Moussallam,
  %``The K*0 (800) scalar resonance from Roy-Steiner representations of pi K scattering,''
  Eur.\ Phys.\ J.\  {\bf C48}, 553 (2006).
%  [hep-ph/0607133].


%\cite{Truong:1988zp}
\bibitem{IAM}
T.~N.~Truong,
%``Chiral Perturbation Theory And Final State Theorem,''
Phys.\ Rev.\ Lett.\  {\bf 61} (1988) 2526.
%%CITATION = PRLTA,61,2526;%%
Phys.\ Rev.\ Lett.\ {\bf 67}, (1991) 2260;
A. Dobado et al., Phys.\ Lett.\ {\bf B235} (1990) 134.
%\bibitem{Dobado:1992ha}
%\cite{Dobado:1996ps}
%\bibitem{Dobado:1996ps}
A.~Dobado and J.~R.~Pel\'aez,
%``A Global fit of pi pi and pi K elastic scattering in 
%ChPT with dispersion relations,''
Phys.\ Rev.\ D {\bf 47} (1993) 4883;
%[arXiv:hep-ph/9301276].
%%CITATION = HEP-PH 9301276;%%
%``The inverse amplitude method in Chiral Perturbation Theory,''
Phys.\ Rev.\ D {\bf 56} (1997) 3057.
%[arXiv:hep-ph/9604416].
%%CITATION = HEP-PH 9604416;%%
%\cite{Dobado:1992ha}

%\cite{LlanesEstrada:2011kz}
\bibitem{LlanesEstrada:2011kz}
  F.~J.~Llanes-Estrada, J.~R.~Pelaez, J.~Ruiz de Elvira,
  %``Fock space expansion of sigma meson in leading-Nc,''
  Nucl.\ Phys.\ Proc.\ Suppl.\  {\bf 207-208}, 169-172 (2010).
  [arXiv:1101.2539 [hep-ph]].


\bibitem{Jaffe} R. L. Jaffe, Proceedings of the Intl. Symposium
on Lepton and Photon Interactions at High Energies. Physikalisches Institut, University of Bonn (1981) . ISBN: 3-9800625-0-3 


%\cite{Espriu:1989ff}
\bibitem{chptlargen}
%\cite{Andrianov:ay}
%\bibitem{Andrianov:ay}
A.~A.~Andrianov,
%``Bosonization In Four-Dimensions Due To Anomalies And An Effective Lagrangian For Pseudoscalar Mesons,''
Phys.\ Lett.\ B {\bf 157}, 425 (1985).
%%CITATION = PHLTA,B157,425;%%
%\cite{Andrianov:1983fg}
%\bibitem{Andrianov:1983fg}
A.~A.~Andrianov and L.~Bonora,
%``Finite - Mode Regularization Of The Fermion Functional Integral,''
Nucl.\ Phys.\ B {\bf 233}, 232 (1984).
%%CITATION = NUPHA,B233,232;%%
D.~Espriu, E.~de Rafael and J.~Taron,
%``The QCD Effective Action At Long Distances,''
Nucl.\ Phys.\ B {\bf 345} (1990) 22
%[Erratum-ibid.\ B {\bf 355} (1991) 278].
%%CITATION = NUPHA,B345,22;%%
%\cite{Peris:1994dh}
%\bibitem{Peris:1994dh}
S.~Peris and E.~de Rafael,
%``On the large N(c) behavior of the L(7) coupling in chi(PT),''
Phys.\ Lett.\ B {\bf 348} (1995) 539
%[arXiv:hep-ph/9412343].
%%CITATION = HEP-PH 9412343;%%



%\cite{Nebreda:2011di}
\bibitem{Nebreda:2011di}
  J.~Nebreda, J.~R.~Pelaez and G.~Rios,
  %``Chiral extrapolation of pion-pion scattering phase shifts within standard
  %and unitarized Chiral Perturbation Theory,''
  Phys.\ Rev.\  D {\bf 83}, 094011 (2011).
%  [arXiv:1101.2171 [hep-ph]].
  %%CITATION = PHRVA,D83,094011;%%


%\cite{Pelaez:2010fj}
\bibitem{Pelaez:2010fj}
  J.~R.~Pelaez and G.~Rios,
  %``Chiral extrapolation of light resonances from one and two-loop unitarized
  %Chiral Perturbation Theory versus lattice results,''
  Phys.\ Rev.\  D {\bf 82}, 114002 (2010).
%  [arXiv:1010.6008 [hep-ph]].
  %%CITATION = PHRVA,D82,114002;%%




%\cite{Nebreda:2010wv}
\bibitem{Nebreda:2010wv}
  J.~Nebreda and J.~R.~Pelaez.,
  %``Strange and non-strange quark mass dependence of elastic light resonances
  %from SU(3) Unitarized Chiral Perturbation Theory to one loop,''
  Phys.\ Rev.\  D {\bf 81}, 054035 (2010)
%  [arXiv:1001.5237 [hep-ph]].
  %%CITATION = PHRVA,D81,054035;%%


%\cite{GomezNicola:2001as}
\bibitem{GomezNicola:2001as}
  A.~Gomez Nicola and J.~R.~Pelaez,
  %``Meson meson scattering within one loop chiral perturbation theory and  its
  %unitarization,''
  Phys.\ Rev.\  D {\bf 65}, 054009 (2002).
%  [arXiv:hep-ph/0109056].
  %%CITATION = PHRVA,D65,054009;%%



\end{thebibliography}
\end{document}